\documentclass[aip
 amsmath,amssymb,
 reprint,%
]{revtex4-1}
\usepackage[colorlinks=true,citecolor=black,
    linkcolor=black, urlcolor=black]{hyperref}

\usepackage{graphicx}% Include figure files
\usepackage{caption}
\usepackage{subcaption}
\usepackage{dcolumn}% Align table columns on decimal point
\usepackage{bm}% bold math
\usepackage{amsmath}
\usepackage[utf8]{inputenc}
\usepackage[T1]{fontenc}
\usepackage{mathptmx}
\usepackage{etoolbox}
\usepackage{xcolor}
\definecolor{mplC0}{HTML}{1F77B4}
\definecolor{mplC1}{HTML}{FF7F0E}
\definecolor{mplC2}{HTML}{2CA02C}
\definecolor{seabornRed}{HTML}{FF0000}
\definecolor{seabornTeal}{HTML}{008080}
\definecolor{seabornPurple}{HTML}{800080}

\makeatletter
\def\@email#1#2{%
 \endgroup
 \patchcmd{\titleblock@produce}
  {\frontmatter@RRAPformat}
  {\frontmatter@RRAPformat{\produce@RRAP{*#1\href{mailto:#2}{#2}}}\frontmatter@RRAPformat}
  {}{}
}%
\makeatother
\begin{document}

\preprint{AIP/123-QED}

\title[3D Integrated Embedded Filters for Superconducting Quantum Circuits]{3D Integrated Embedded Filters for Superconducting Quantum Circuits}
\author{W. Ahmad}
 \affiliation{Oxford Quantum Circuits Ltd}%Lines break automatically or can be forced with \\
\author{G. Consani}
 \affiliation{Oxford Quantum Circuits Ltd}
\author{M. T. Haque}
 \affiliation{Oxford Quantum Circuits Ltd}
\author{J. Dunstan}
 \affiliation{Oxford Quantum Circuits Ltd}
\author{B. Vlastakis}%
 % \email{bvlastakis@oqc.tech}
\affiliation{Oxford Quantum Circuits Ltd}%

\date{\today}% It is always \today, today,

\begin{abstract}

Microwave filtering for superconducting qubits is a key element of quantum computing technology, enabling high coherence and fast state detection. This work presents the design and implementation of novel microwave Purcell filters for superconducting quantum circuits, integrated within a multilayer printed circuit board (PCB). The off-chip design removes all filter components from the qubit substrate, reducing device complexity, improving layout footprint and allowing better scalability to large qubit counts. Each embedded filter can couple up to nine readout resonators, enabling efficient multiplexed readout. Electromagnetic simulations of the filter predict a thousand-fold improvement in qubit isolation from the readout port. The design was experimentally validated under cryogenic conditions in conjunction with a 35-qubit device, demonstrating compatibility of the PCB-based filter with high-coherence superconducting qubits. The comparison of the measured qubit median $T_1$ of 84~\textmu s with the expected radiative limit from electromagnetic simulations validated the presence of Purcell filtering in the system.

\end{abstract}

\maketitle

\section{\label{sec:introduction} Introduction}
% Quantum Computation requires filtering
Quantum computation requires the accurate preparation, manipulation and readout of quantum information \cite{divincenzo_topics_1997}.
While multiple architectures provide paths for realizing quantum computation, superconducting qubits are a leading platform due to their scalability and fine-tuned control \cite{devoret_superconducting_2013,kim_evidence_2023,acharya_integration_2024}.
At the same time, superconducting qubits are sensitive to decoherence, including that caused by energy relaxation.
Compounding to this, fast, high fidelity and non-destructive readout is important to an architecture’s viability for error-corrected operation.
Typical readout methods in superconducting circuits introduce a trade-off between the readout speed and the qubit coherence.
This effect must be controlled in a scalable way for fault-tolerant quantum computation \cite{nielsen_quantum_2010}.

%Move trade-off description above the 'solution', the Purcell filter.
Fundamentally, the qubit energy relaxation rate $\Gamma_1(\omega_{qb})$ is controlled by the dissipative environment at the qubit transition frequency $\omega_{qb}$.  In the context of circuit QED, readout of a superconducting qubit with a linear resonator at a frequency $\omega_{res}$, dispersively-coupled with a coupling rate $g$, results in a trade-off between the external coupling rate of the resonator $\kappa_\textrm{ext}$ (and therefore the readout speed), and the qubit relaxation rate $\Gamma_1(\omega_{qb}) \geq (g/(\omega_{qb}-\omega_{res}))^2\kappa_{\textrm{ext}}(\omega_{qb})$ \cite{krantz_quantum_2019, sete_quantum_2015}.

% Purcell Filtering Description
Purcell filtering is a way to address this tradeoff. By incorporating microwave filtering (i.e. a stopband) in close proximity to a superconducting qubit, the dissipative environment can be engineered to both protect the qubit energy relaxation and enable a large external coupling rate of the resonator, and therefore fast readout \cite{girvin_circuit_2009}.

%Methods of filtering to-date
To date, numerous Purcell filters have been realized in superconducting circuits, typically using transmission line-based bandstop and bandpass filter designs \cite{reed_fast_2010, sete_quantum_2015}. 
Such designs have been augmented to incorporate multiplexed qubit readout \cite{jeffrey_fast_2014,heinsoo_rapid_2018}, multi-mode protection \cite{yan_broadband_2023, park_characterization_2024}, quarter-wavelength and half-wavelength on-chip filters \cite{ref_s42005-024-01733-3}, and intrinsic filtering \cite{bronn_reducing_2015,sunada_fast_2022,spring_high_2022}.
Due to the single-layer layout and subsequent signal routing,  the filter footprint is often larger than the qubit itself, leading to an increase in the size of the  superconducting quantum processing unit (QPU) or a compromise in the readout resonator bandwidth~\cite{acharya_quantum_2024}.
To address this challenge, filter miniaturization and multilayer signal integration \cite{bronn_broadband_2015, hazra_benchmarking_2024, bakr_multiplexed_2024} have been demonstrated. However, these examples require additional fabrication and packaging processes which can increase the complexity and the duration of QPU manufacturing.

While many superconducting QPU architectures use printed circuit board (PCB) packaging geared towards signal delivery, to-date there has been no demonstration of the integration of Purcell filters into such packaging. 

In this paper, Purcell filtering embedded in a multilayer superconducting QPU is demonstrated. The proposed design forms a bandpass filter, centered at a desired readout frequency band, from a PCB-integrated multilayer patch antenna.
The off-chip design removes all filter components from the qubit substrate itself. This allows to partially separate the development of the superconducting stack (qubits and resonators), from that of the readout and filtering stack. The latter can therefore evolve to include more complex and cascaded out-of plane filtering solutions, while maintaining a modular and tileable structure. 
Each embedded filter presented here can couple to up to nine readout resonators simultaneously, allowing multiplexed readout. The filter requires no planar signal routing and, even with the output RF connector included, fits within the nine-qubit footprint - enabling scaling to  QPUs of an arbitrary size.

\begin{figure*}[thbp]
    \begin{subfigure}[b]{0.28\textwidth}
        \centering
        \phantomcaption
        \includegraphics[width=\textwidth]{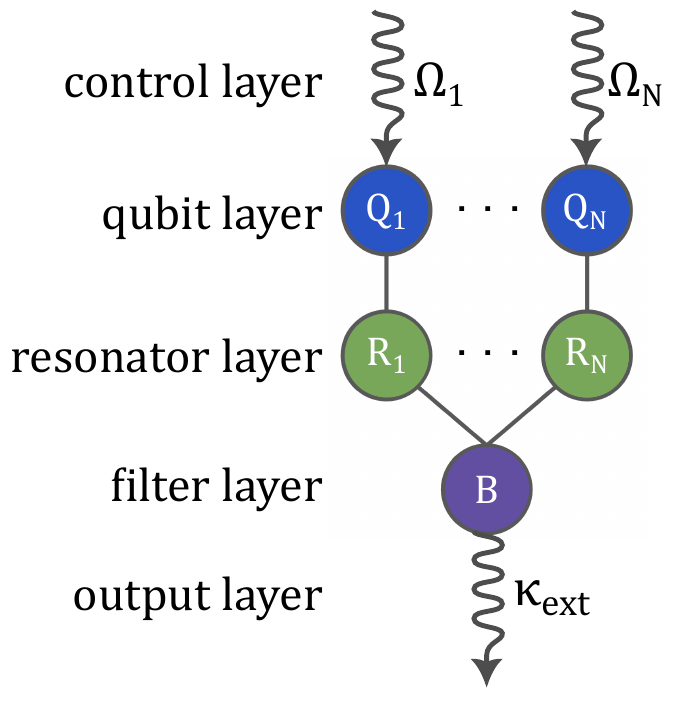}
        \put(-145,141){\small\textbf{(a)}}
%        \caption{}
        \label{fig1:sub1}
    \end{subfigure}
    \hspace{0.035\textwidth}
    \begin{subfigure}[b]{0.55\textwidth}
        \centering
        \phantomcaption
        \includegraphics[width=\textwidth]{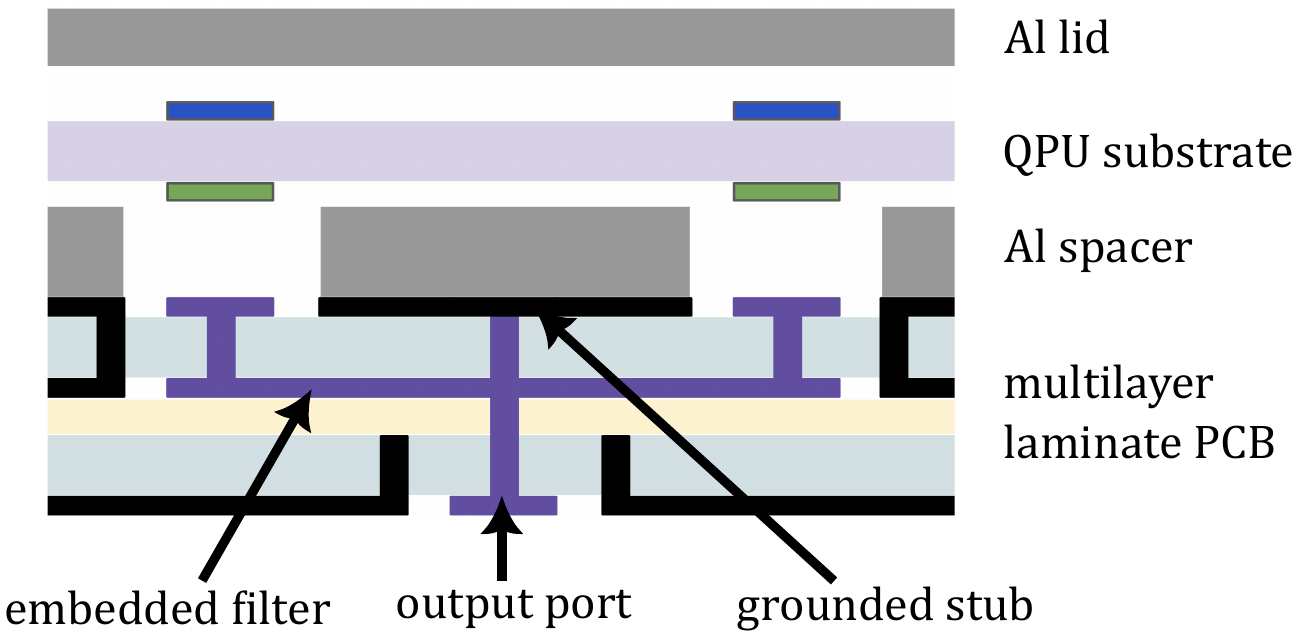}
%        \caption{}
        \put(-277,143){\small\textbf{(b)}}
        \label{fig1:sub2}
    \end{subfigure}
    \caption{\textbf{Processor readout stack}. \textbf{a.} A diagram representing the components of a vertically-integrated QPU. This includes qubits $\textrm{Q}_1$ to $\textrm{Q}_\textrm{N}$ (blue), with individual control lines (with associated drive rates $\Omega_1$ to $\Omega_\textrm{N}$) and individually coupled readout resonators ($\textrm{R}_1$ to $\textrm{R}_\textrm{N}$, in green). The resonators interface with multiplexed bandpass filter(s) (B, in purple) whose coupling rate to the output line is denoted by $\kappa_\mathrm{ext}$. \textbf{b.} An illustration of the cross-sectional view of the physical layout of the readout stack (with no dedicated qubit control side); showing the qubit chip coupled to a unit cell of the PCB via an aluminum spacer.}
    \label{fig1}
\end{figure*}

In the work presented here, simulations of the PCB-based filter, in association with the QPU, were used to determine a thousand-fold protection of the qubits from the radiative decay channel into the readout port, compared to a readout solution without any filtering. A physical realization of the filter was then characterized with high-coherence superconducting qubits \cite{acharya_integration_2024}, demonstrating that the filters are compatible with high coherence ($T_1$ and $T_{2,echo}>$ 80~\textmu s), without sacrificing the readout resonator speed. Comparing simulations with the experimentally measured qubit relaxation times confirmed the presence of Purcell filtering in action.

The paper is organized in four main sections. Section II covers the design and implementation of the building block unit cell and the full PCB stack. Section III details the simulation setup of the PCB, both stand-alone and interfaced with a QPU. Sections IV and V cover the measurement results and the conclusions drawn from the work undertaken.

\section{Design and Implementation}

%   1. Overview, Unit cell description, interface w/ QPU, multilayer, 
%   2. Figure 1; Processor Readout Stack

The vertically-integrated design philosophy splits a QPU into layers, containing qubits, resonators, filters, control and readout (see fig. \ref{fig1}a). This architecture can be realized with a system of qubits and resonators sharing the same substrate, using double-sided fabrication~\cite{rahamim_double-sided_2017}, and a multilayer PCB containing filtering and readout connectorization. For the purpose of this work, a simplified architecture without a dedicated individual qubit control layer is considered (see fig. \ref{fig1}b). 
Such a stack forms a unit cell for QPU design, which can be tiled to work with large-scale multi-qubit QPUs.
In this section, a 9--1 embedded filter design is reviewed, it is shown how the design can be tiled to an arbitrary processor size, and its incorporation into a physical device.

\subsection{Design of a Single Unit Cell with 9--1 Multiplexing and Filtering}

\begin{figure*}
    \centering
    \begin{subfigure}[b]{0.4\textwidth}
        \centering
        \phantomcaption
        \includegraphics[width=\textwidth]{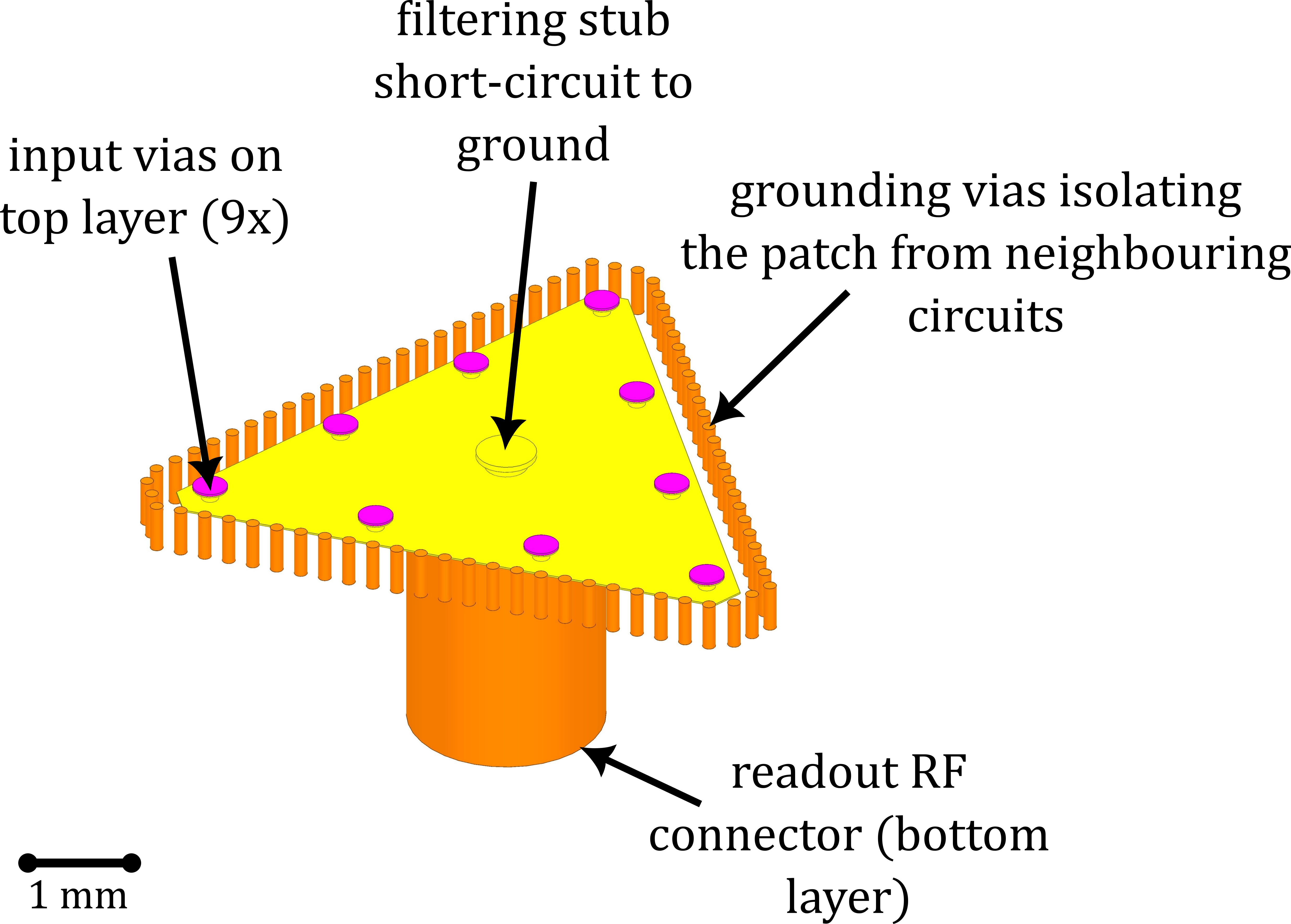}
        \put(-210,150){\small\textbf{(a)}}
        \label{fig2:sub1}
    \end{subfigure}
    \hspace{0.03\textwidth}
    \begin{subfigure}[b]{0.3\textwidth}
        \centering
        \phantomcaption
        \includegraphics[width=\textwidth]{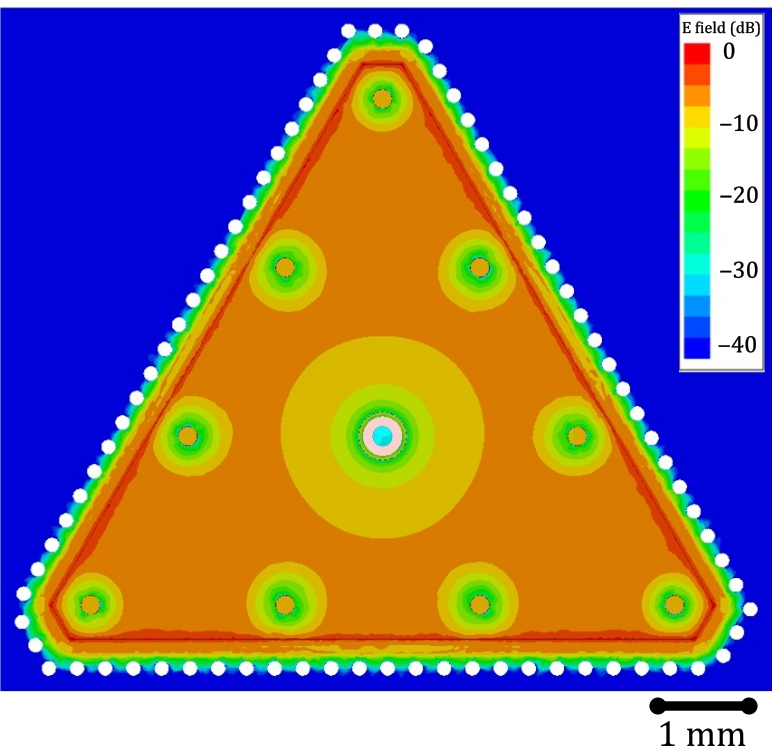}
        \put(-170,150){\small\textbf{(b)}}
        \label{fig2:sub2}
    \end{subfigure}
    \caption{\textbf{Embedded filter}. \textbf{a.} Isometric top view of the building block 9--1 unit cell; showing the input vias, the embedded filter and the output port. Each filter is a triangular shape whose fundamental resonant mode determines the passband center frequency. Each filter capacitively couples to up to a maximum of nine readout resonators. A short-circuited stub in the center of the filter determines the filter bandwidth. \textbf{b.} Electric field profile from an eigenmode simulation of the embedded filter, corresponding to the mode at the fundamental frequency of the filter. This mode follows a drum-head behavior with a maximum field at the outside edges and a maximum current at the center. This enables a strong capacitive coupling to the readout resonators, while allowing a strong current coupling to a low impedance output port.}
    \label{fig2}
\end{figure*}

The design of a single filter unit cell acts as the basic building block for the rest of the PCB package. This is shown in fig. \ref{fig2}a. For the demonstration of this concept, a filter which protects up to nine qubits and enables 9--1 frequency multiplexed readout is selected. The PCB stack comprises three layers.

%Resonant structure (middle layer)
Embedded in the middle layer, the filter takes the form of a triangular-shaped coplanar patch antenna, with an out-of-plane feedline at its center. The size of the patch is designed to have a fundamental resonant mode that matches a selected readout center frequency of 9.8 GHz, using a variation of the triangular patch equation from \cite{ref_antenna_handbook_garg&bahl, ref_antenna_handbook_lo&lee} given below in eq. \ref{eq:patch},
\begin{equation}
a = \rho \frac {{2}{c}} {{3}{f_r}\sqrt{{\epsilon_r}}},
\label{eq:patch}
\end{equation}
where $a$ is the outer side of the patch, $c$ is the speed of light, $f_r$ is the center frequency of the bandpass filter, $\epsilon_r$ is the dielectric permittivity of the patch substrate and $\rho$ is a pre-factor dependent on the relative participation of the shorting stub inductance in the center of the stack. The shape is chosen in order to maximize the coverage of qubits while maintaining symmetry for tiling along the lattice and the most attainable bandwidth. For the designed triangular shape, the value of $\rho$ is $\sim\frac{1}{2}$.

% Input interface (top layer)
On the top layer of the PCB, there are nine vias, which act as the nine input ports, capacitively coupling to nine resonators in the QPU. The locations of these vias are designed to match the layout of the qubit chip.
% and due to the shape of the patch itself, the nine vias are not equidistant from the center of the patch: six vias are nearer the center, whilst the remaining three fall near the three corners of the patch. This can cause a variation in the resonator-filter capacitive coupling. However, this can be easily mitigated by modifying the via geometry for any one set of the affected vias.
% Output interface (bottom layer)
The bandpass filtering effect within the unit cell is achieved by integrating a via which runs in the center of the patch and connects all three layers of the PCB stack (see fig. \ref{fig1}b). On the top layer, the via acts as a grounded inductive shunting stub, whose size controls the frequency and the bandwidth of the ensuing filter. On the bottom layer, the via is directly connected to a 50 $\Omega$ output port, in the form of a surface-mount SMPS connector. The use of a patch for this design, as opposed to traditional transmission line tracks, allows it to attain a relatively wide filter bandwidth (> 500 MHz).

The filter stack is designed so that the filtering patch in the middle layer is shielded by two ground planes, and that each filter is surrounded by grounded shielding vias to further minimize crosstalk with neighboring filters (see fig. \ref{fig2}a).

\subsection{Implementation of the 35--6 Multiplexing and Filtering PCB Stack}

The PCB package is designed to be compatible with an OQC-Toshiko QPU~\cite{acharya_integration_2024}. This PCB multiplexes 35 resonators, connecting to six RF output lines. The three-layer PCB is constructed of two laminated substrates bound by a pre-impregnated fiberglass composite (pre-preg).
The two dielectric substrates are Rogers RT/duroid 5880, having a dielectric permittivity $\epsilon_r = 2.2$ and a loss tangent tan$\delta$ = $9\times10^{-4}$ ~\cite{ref_Rogers_dielectric}. The bottom substrate has a thickness of 0.254 mm and the top substrate has a thickness of 0.127 mm. Rogers Bondply 2929 is then used as a pre-preg to bond the two substrates with each other. The pre-preg has a dielectric permittivity of $\epsilon_r = 2.94$, a tan$\delta$ of $3\times10^{-3}$ and a thickness of 0.076 mm~\cite{ref_Rogers_pre-preg}. The stack consists of three copper layers (conductivity $\sigma = 5.8\times10^7$ S/m and thickness of $18$ \textmu m), where the outside surfaces of the two outer layers are finished with silver immersion plating (conductivity $\sigma = 6.3\times10^7$ S/m and thickness of 0.2--0.3 \textmu m). 

\begin{figure*}
    \centering
    \begin{subfigure}[b]{0.25\textwidth}
        \centering
        \phantomcaption
        \includegraphics[width=\textwidth]{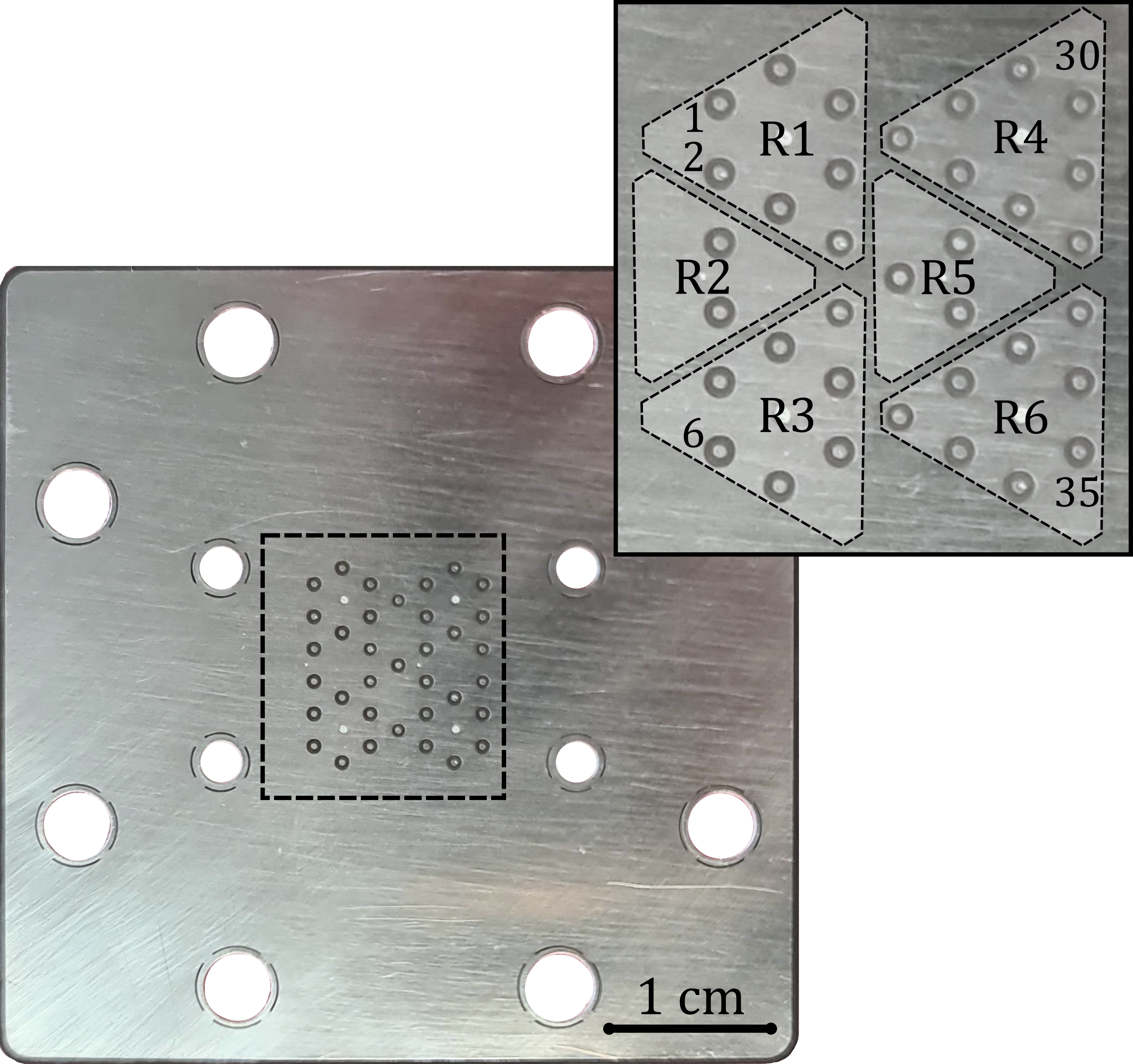}
        \put(-125,98){\small\textbf{(a)}}
        \label{fig3:sub1}
    \end{subfigure}
    \hfill
    \begin{subfigure}[b]{0.3\textwidth}
        \centering
        \phantomcaption
        \includegraphics[width=\textwidth]{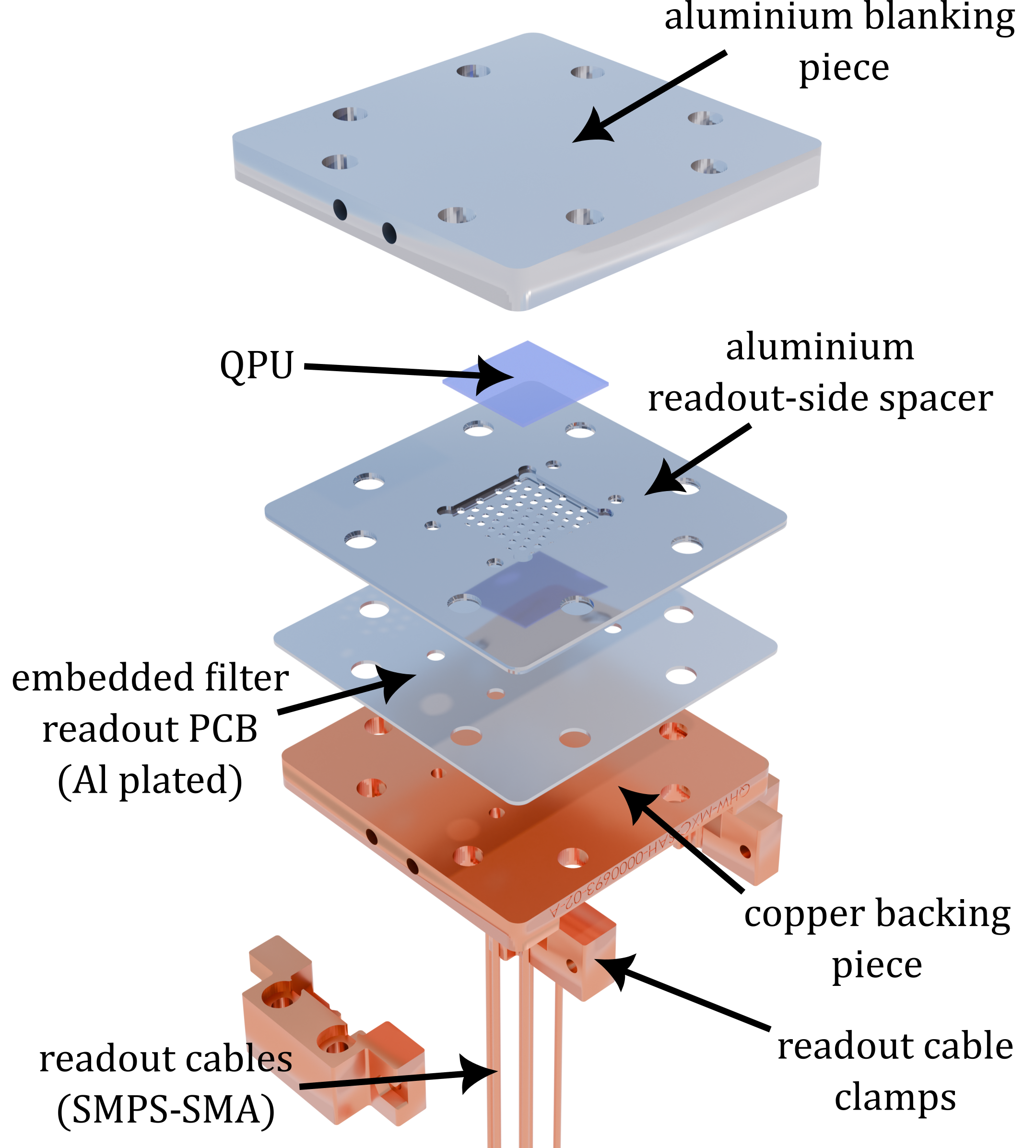}
        \put(-145,160){\small\textbf{(b)}}
        \label{fig3:sub2}
    \end{subfigure}
    \hfill
    \begin{subfigure}[b]{0.4\textwidth}
        \centering
        \phantomcaption
        \includegraphics[width=\textwidth]{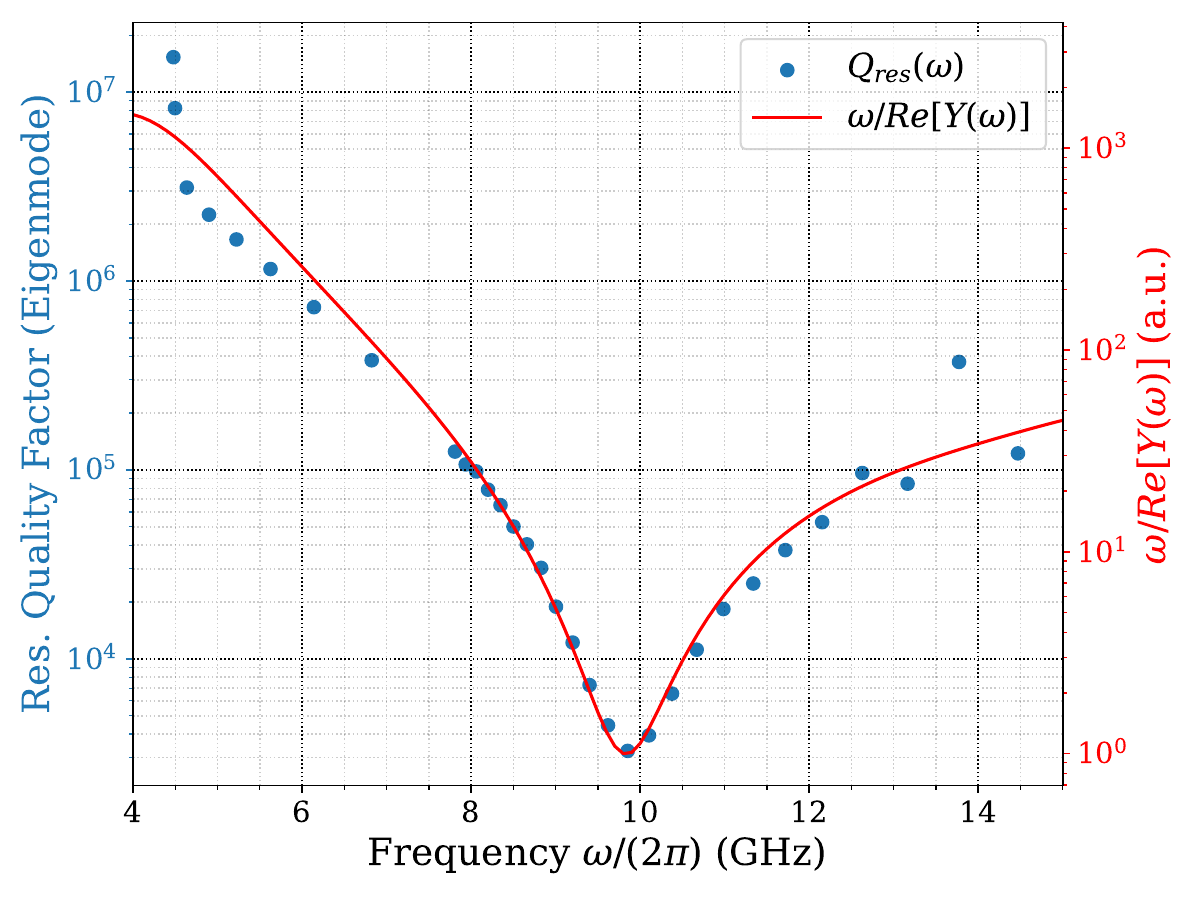}
        \put(-200,155){\small\textbf{(c)}}
        \label{fig3:sub3}
    \end{subfigure}
    \caption{\textbf{Readout PCB}.
    \textbf{a.} Photograph of the top (readout-resonator facing) layer of the readout PCB for at 35-qubit processor. The inset shows a zoomed version of the area within the dashed rectangle. Here one can see how the input vias (numbered 1 to 35) are arranged in a triangular grid, matching that of the qubits and readout resonators in the QPU. The embedded filters are tiled to enable full processor readout with six filters (highlighted within dashed lines and numbered R1 to R6 in the inset image).
    \textbf{b.} 3D rendering image of the physical QPU assembly, including the QPU itself, the embedded filter PCB, the readout wiring, and metallic sample holder parts. \textbf{c.} Blue scatter data (left vertical axis): readout resonator external quality factors, estimated from a full electromagnetic eigenmode simulation, including the PCB with embedded filter, nine resonators and nine qubits. Solid red line (right vertical axis): normalized quality factor inferred from a modal network simulation of the standalone PCB.}
    \label{fig3}
\end{figure*}

The physically-implemented PCB comprises six tiled filters, with the layout shown in fig. \ref{fig3}a. Four of these filters are 9--1 designs described previously, and the two remaining ones are 3--1 variants. While the PCB stack can support connections to 42 resonators, only 35 are used in this implementation.

\section{Modeling and Simulation}

Simulations are performed to understand the electromagnetic behavior of the embedded filters in the PCB and the PCB coupling interaction with a QPU. Ansys EDT~\cite{ref_Ansys_HFSS} is used for these electromagnetic finite element simulations. The simulations are conducted in two stages: first with the stand-alone PCB, and then with the PCB coupled to a system of coaxial qubits and resonators, on two sides of a shared substrate. The first stage is aimed at checking the properties of the PCB on its own, and ensuring  that the multiplexing and filter center frequency and the bandwidth are correct. The second stage consists of electromagnetic simulations of a combined model of the PCB and the qubit processor to more precisely determine the quality factors of the resonators and the qubits, and to assess the filtering nature of the PCB. In order to make the simulations less computationally intensive, only one unit cell of the stack is simulated, consisting of an embedded filter with 9--1 multiplexing, and the electrodes of the corresponding nine readout resonators and nine qubits. Due to the electromagnetic isolation between the filters, a negligible tradeoff in the accuracy of the simulations is assumed. Furthermore, losses in the metal conductor layers and dielectric losses are not included in the models to minimize the runtime of the simulations. These will be investigated in future work (also see~\cite{kennedy2026designoperationwaferscalepackages}).

\subsection{Modal Analysis of the PCB in Isolation}

In order to check the stand-alone PCB in the first stage of the simulations, the center frequency and the bandwidth of the filter are characterized via modal network simulations at the different resonator port locations. This involves analyzing the admittance at the weakly-coupled input ports against frequency. Such a simulation enables a relatively fast prediction of the response of the resonator, as a function of its detuning from the embedded filter \cite{1003.0142}. Modeling the response of the resonator to the embedded filter produces the relationship (see \cite{bronn_reducing_2015,sunada_fast_2022,spring_high_2022})
\begin{equation}
Q = \frac{\omega_r C}{Re[Y(\omega_r)]},
\label{eq:QofY}
\end{equation}
where in this case $Q$ is the quality factor of the resonator, $\omega_r$ is the angular frequency of the resonator, $C$ is the resonator shunt capacitance and $\textrm{Re}[Y(\omega)]$ is the real part of the input admittance. When all the sources of loss in the materials (i.e. conductive and dielectric losses) are removed from the simulations, as is the case in the present work, $Q$ can be interpreted as the resonator external quality factor $Q_{ext}$, which only depends on the coupling of the resonator to its readout line. Unless explicitly stated otherwise, all quality factors given below are external quality factors ($Q=Q_{ext}$). As noted in \cite{1003.0142, 0803.4490v1}, eq. \ref{eq:QofY} closely models the observed external quality factor when all the coupling modes of the surrounding cavity are taken into account in the calculation of the admittance.

So as to implement this simulation setup in the electromagnetic model, the input vias are overlaid with circular pads (of the same diameter as the vias). A capacitive boundary condition is set on these pads to mimic the coupling gap between the actual readout resonators in the QPU and the PCB. This capacitance is estimated to be 1.73 fF from independent electromagnetic simulations. These pads are then overlaid with perfect electrical conducting pads which act as 50 $\Omega$ wave ports for the nine signal inputs. Since 50 $\Omega$ RF connectors are used at the output of the PCB, an RLC boundary condition, with a resistance of 50 $\Omega$, is set on the dielectric of the output via. The model is then simulated over a relevant frequency range of 4--15 GHz.

Using the simulation results of the modal network analysis~\cite{ref_Ansys_HFSS}, the quality factor from eq. \ref{eq:QofY} is estimated over a range of frequencies, resulting in a filtering response curve, showing the center frequency and the bandwidth of the Purcell filter. The result is presented in figure \ref{fig3}c, as the solid red curve (values on the right vertical axis). Since what is most important is the relative value of the filter quality factor, rather than its absolute value at each frequency, all the results of modal network simulations are displayed in terms of $Q(\omega)/C=\omega/Re[Y(\omega)]$. In this case, the (modal network simulation) data in the figure has been further normalized to its minimum value, $\min_\omega(\omega/Re[Y(\omega)])$, which occurs at the filter center frequency. The clear dip at center of the designed passband, at 9.8 GHz, demonstrates that the PCB is working as a bandpass filter. This is the simulated center frequency of the Purcell filter. The 3 dB bandwidth of the passband is determined to be 0.9 GHz. The filtering ratio, i.e. the ratio of the attenuation or the Q factors at 4.4 GHz, the approximate qubit frequency, to 9.8 GHz, the readout resonator frequency, is $\sim31$ dB.

The blue scatter data (\textcolor{mplC0}{\scalebox{1.1}{$\bullet$}}) in fig. \ref{fig3}c represents the absolute quality factor of one of the readout resonators, obtained from an eigenmode simulation of a model that includes the resonators and the qubits, as detailed in the next subsection. Note the good agreement of the full eigenmode simulation with the modal network simulation.

\subsection{Eigenmode Analysis of the PCB with a QPU}
\begin{figure*}[!bhtp]
    \centering
    \begin{subfigure}[b]{0.49\textwidth}
        \centering
        \phantomcaption
        \includegraphics[width=\textwidth]{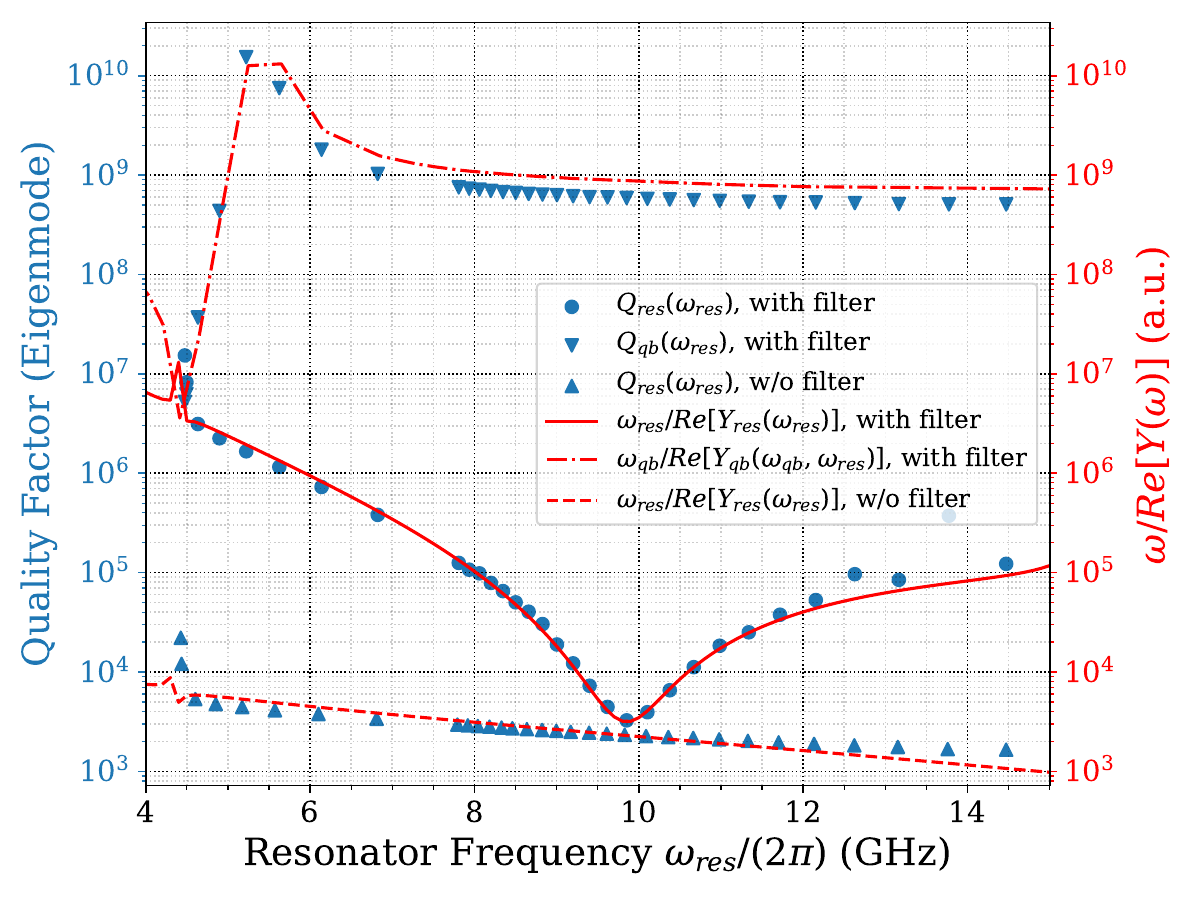}
        \put(-240,190){\small\textbf{(a)}}
        \label{fig4:sub1}
    \end{subfigure}
    \hfill
    \begin{subfigure}[b]{0.49\textwidth}
        \centering
        \phantomcaption
        \includegraphics[width=\textwidth]{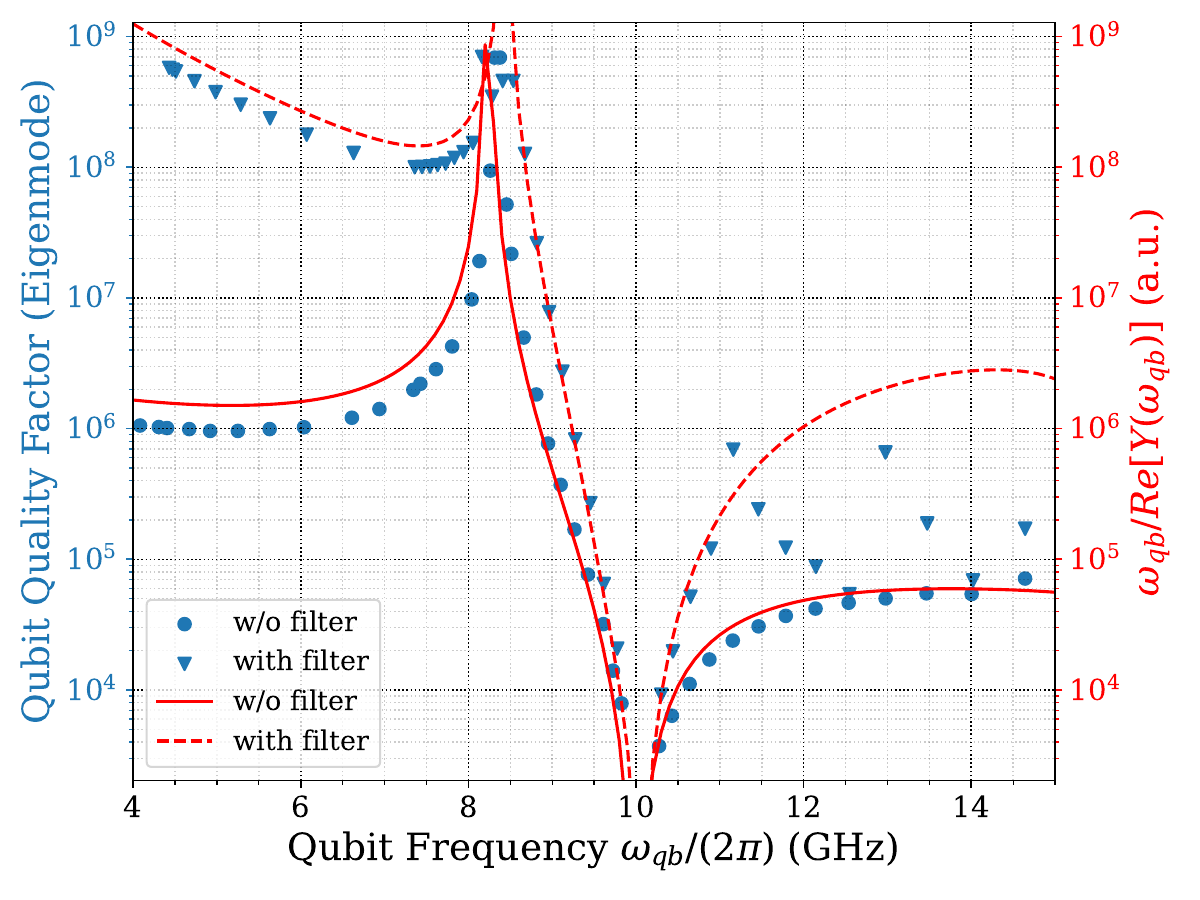}
        \put(-240,190){\small\textbf{(b)}}
        \label{fig4:sub2}
    \end{subfigure}
    \caption{\textbf{Filter finite-element modeling}. \textbf{a.} Quality factors of the qubit and the resonator modes, as a function of varying resonator frequency, extracted from eigenmode (blue scatter data, left vertical axis) and modal network (red lines, right vertical axis) finite-element simulations. \textbf{b.} Comparison of the qubit mode quality factors in a packaged QPU system with (\textcolor{mplC0}{\scalebox{0.9}{$\blacktriangledown$}} data and red dashed line) and without (\textcolor{mplC0}{\scalebox{1.1}{$\bullet$}} data and red solid line) embedded PCB filter.}
    \label{fig4}
\end{figure*}

For the second phase of the simulations, the model from the first simulation setup is taken as the starting point. The capacitor pads (which originally emulated the capacitive coupling to the resonators) and the wave port pads are removed from the top of the input vias. On the output side, the RLC boundary condition of 50 $\Omega$ resistance is maintained across the dielectric of the output via to retain the conditions of 50 $\Omega$ RF connectors. A 260 \textmu m-thick spacer, in the form of a perfect electrical conductor object, and having nine holes, aligned to the nine input vias of the PCB unit cell, is placed on top of the PCB. This spacer is a precise model of the aluminum spacer in the physical setup of the sample holder assembly (see fig. \ref{fig3}b), and defines the geometry of the coupling environment within which the photon transmission between the qubit chip and the PCB takes place. The diameter of each of the holes is 1.1 mm. This is set in consideration of the size of the outer pad of the readout resonators (1 mm maximum diameter) and to stop any short-circuit between the pads of the resonators and the metal spacer itself. The aluminum spacer is then overlaid with a model of a nine-qubit section of the QPU chip. The chip substrate is modeled as cryogenic-sapphire, an anisotropic material with a dielectric permittivity of $(\epsilon_{rx}, \epsilon_{ry}, \epsilon_{rz}) = (11.4, 11.4, 9.26)$~\cite{ref_PhysRevApplied.19.034064-accepted, ref_Krupka1999WGM, ref_shelby1980}; the qubits are laid on the top side and the readout resonators on the bottom side, facing the PCB filter~\cite{acharya_integration_2024}. 

An RLC boundary condition, set to inductance, is then applied to the lead connecting the two capacitive pads of one of the nine resonators. This inductance replaces the geometric inductance of the spiral connecting the two concentric electrodes of the physical resonator, and it is set to 1.8 nH to give a resonator mode at 9.8 GHz. The same operation is performed on one of the qubits, the one matching the location of the shunted resonator. In this case, the inductance value is set to 11.5 nH, resulting in a (linear) qubit mode frequency of $\sim4.43$ GHz. The model is then setup with eigenmode parameters and the design is simulated.

The relevant eigenmode results are represented as blue scatter points in fig. \ref{fig4}a, whose values can be read on the left vertical axis. Naturally, the system displays three fundamental resonant modes: one corresponding to the embedded filter, which is identified as having the lowest quality factor, one corresponding to the qubit, at a frequency of $\sim4.43$ GHz, and one corresponding to the resonator.

Figure \ref{fig2}b. shows the electric field distribution associated with the filter mode (mode frequency of 9.8 GHz) at the top surface of the embedded filter PCB. It can be seen that the field is spread throughout the patch, but more concentrated towards the outer sides of the patch. There is minimum coupling from the patch to the surrounding region outside; which is an indication of the minimum crosstalk between neighboring filters. Additionally, at the center frequency of the filter, the region occupied by the nine resonators shows minimum field concentration indicating that this is indeed the filter mode.

By sweeping the value of the inductance of the lumped inductor defined between the resonator pads, one can see the frequency of the resonator mode span between 4 GHz and 15 GHz. Then it becomes straightforward to identify the resonator mode and its quality factor. The magnitude of this quality factor is intermediate between that of the filter, below (at a value of $\sim10$, not shown in figure \ref{fig4}a), and that of the qubit, above. It also displays the desired frequency dispersion, having a dip at the filter frequency of 9.8 GHz. The corresponding $Q_{res}(\omega_{res})$ data is shown as the blue circles (\textcolor{mplC0}{\scalebox{1.1}{$\bullet$}}) both in fig. \ref{fig3}c and in fig. \ref{fig4}a. Note that the scatter in $Q_{res}(\omega_{res})$ at around $\omega_{res}/(2\pi)\sim13.5$ GHz can be attributed to the fact that the simulation software did not converge with a suitable low error percentage at these values. The solid red line in figure \ref{fig4}a (values on the right vertical axis) represents the quality factor of the resonator (rescaled by the resonator shunt capacitance) $\omega_{res}/Re[Y(\omega_{res})]$, obtained by an equivalent modal network simulation of the system, where the resonator shunting inductor is replaced with a lumped port and $Y(\omega)$ represents the input admittance at this port. Note the nice agreement of this data with the eigenmode data, when represented in an appropriate scale.

Also shown in the same plot are the quality factors of the qubit mode $Q_{qb}(\omega_{res})$ (downward triangles \textcolor{mplC0}{\scalebox{0.9}{$\blacktriangledown$}}). As expected, this displays a weak dependence on the resonator frequency, as long as $\vert\omega_{res}-\omega_{qb}\vert\gg1$. At these frequencies, the qubit mode quality factor is at least four orders of magnitude greater than that of the resonator, due partly to the greater distance between the qubit and the output port, and partly to the effect of the embedded filter (more on this below). As $\omega_{res}$ approaches the frequency of the qubit from above, one first sees an uptick in the qubit mode quality factor, resulting from a destructive interference between the radiative path into the output port mediated by the embedded filter, and the radiative path mediated by both the filter and the resonator (intrinsic Purcell filtering\cite{bronn_reducing_2015,sunada_fast_2022,spring_high_2022}). Then, as $\omega_{res}$ goes through $\omega_{qb}$, $Q_{qb}(\omega_{res})$ decreases (Purcell effect) and $Q_{res}(\omega_{res})$ goes up, as the two modes hybridize. The dot-dashed red line in fig. \ref{fig4}a shows the result of an equivalent modal simulation where the qubit inductor is replaced by a lumped port, which is used to determine the admittance seen by the port, as the resonator frequency is changed, and therefore the qubit normalized quality factor $\omega_{qb}/Re[Y(\omega_{qb},\omega_{res})]$.

Finally, the upright blue triangles \textcolor{mplC0}{\scalebox{0.9}{$\blacktriangle$}} and the dashed red line in fig. \ref{fig4}a represent the quality factor of the resonator mode $Q_{res}(\omega_{res})$, determined from an separate simulation where the PCB filter object has been removed, and 50 $\Omega$ impedance boundaries have been applied directly at the outer edge of the cylindrical holes in the aluminum spacer piece. This configuration represents what would happen if there was a direct capacitive connection between the QPU and the readout lines (cf. ~\cite{acharya_integration_2024,ward2026echo}). As can seen by comparing with the results with the PCB present, the resonator quality factor in this case is uniformly lower, and only presents the $1/\omega_{res}$ scaling expected for a resonator capacitively coupled to a readout line ($Q_{ext}\simeq C/(Z_0C_c^2\omega_{res})$, with $Z_0$ the characteristic impedance of the line, $C$ the resonator shunt capacitance and $C_c$ its coupling capacitance to the line, cf. \cite{kamigaito2020circuit}). As expected, the effect of the embedded filter is to protect modes that are detuned from the filter frequency, by enhancing their quality factor. At the same time, the effect on modes that are within the passband of the filter is very limited, which is key for enabling fast circuit QED readout.

Figure \ref{fig4}b. shows simulation results that support one key result of this paper, namely that our embedded PCB filter offers protection to the qubit mode, allowing for fast readout without compromising on qubit $T_1$ and $T_2$ times. In these simulations, the qubit quality factor is determined, as a function of the frequency of the qubit modes, both in a model with the PCB and embedded filter, and in a model where the PCB is removed and the 50 $\Omega$ lines are capacitively coupled directly to the QPU. The shunting lumped-element inductor of the resonator is now kept fixed in all simulations, resulting in a resonator mode at $\sim9.8$ GHz. In the eigenmode simulations, the size of the qubit shunting inductor is changed, in order to sweep the qubit mode frequency in the range 4--15 GHz, whereas in the modal network simulations a lumped port is defined between the pads of the qubit and used to determine the admittance seen by the qubit, looking into the readout port, and, from it, the (normalized) qubit quality factor.

As can be seen by looking at the eigenmode simulation data (blue scatter points in in fig. \ref{fig4}b, \textcolor{mplC0}{\scalebox{1.1}{$\bullet$}} for the simulation without the filter and \textcolor{mplC0}{\scalebox{0.9}{$\blacktriangledown$}} for the simulation with the filter), at the typical qubit frequency of $\sim4.4$ GHz, the embedded filter suppresses the radiative decay rate ($1/T_{1,rad}=1/(\omega_{qb}Q_{qb,ext})=1/(\omega_{qb}Q_{qb})$) by $\sim28$ dB. As the qubit frequency continues increasing, a region of additionally enhanced protection is encountered at around $\omega_{qb}\sim8.5$ GHz, which results from intrinsic Purcell filtering (and therefore also occurs in the absence of the embedded filter). Then, as the qubit frequency is swept across the common frequency of the readout resonator and the filter, the quality factor is substantially suppressed by the Purcell effect, before it starts to grow again for $\omega_{qb}>9.8$ GHz. Once again, the scatter seen in the eigenmode quality factors at around 12 GHz (see \textcolor{mplC0}{\scalebox{0.9}{$\blacktriangledown$}} data in particular) is likely due to poor convergence of the simulations. The results of the modal network simulations (red lines in fig. \ref{fig4}b) are in good agreement with those of the eigenmode simulations, as can be seen by displaying the normalized quality factor in the appropriate scale (right vertical axis in the figure).
\section{Experimental Results}
In this section, the results of the experimental demonstration of the embedded multiplexer and filter with a superconducting QPU are presented. These measurements were performed with an enclosure variation of the OQC-Toshiko QPU~\cite{acharya_integration_2024}, a processor containing 35 transmon qubits. As shown in fig. \ref{fig3}b, in this case, the enclosure only contains connections to the readout interface, and all microwave control is performed through the readout PCB packaging described in the previous section.
% , by combining control and readout signals at room temperature.

The device was cooled down to 11 mK and tested in a cryogenic environment, following the experimental system outlined in \cite{acharya_integration_2024}. A vector network analyzer was used for continuous-wave analysis and a system of custom RF control cards was employed for time-resolved experiments.

\begin{figure*}[!thbp]
    \centering
    \begin{subfigure}[b]{0.48\textwidth}
        \centering
        \phantomcaption
        \includegraphics[width=\textwidth]{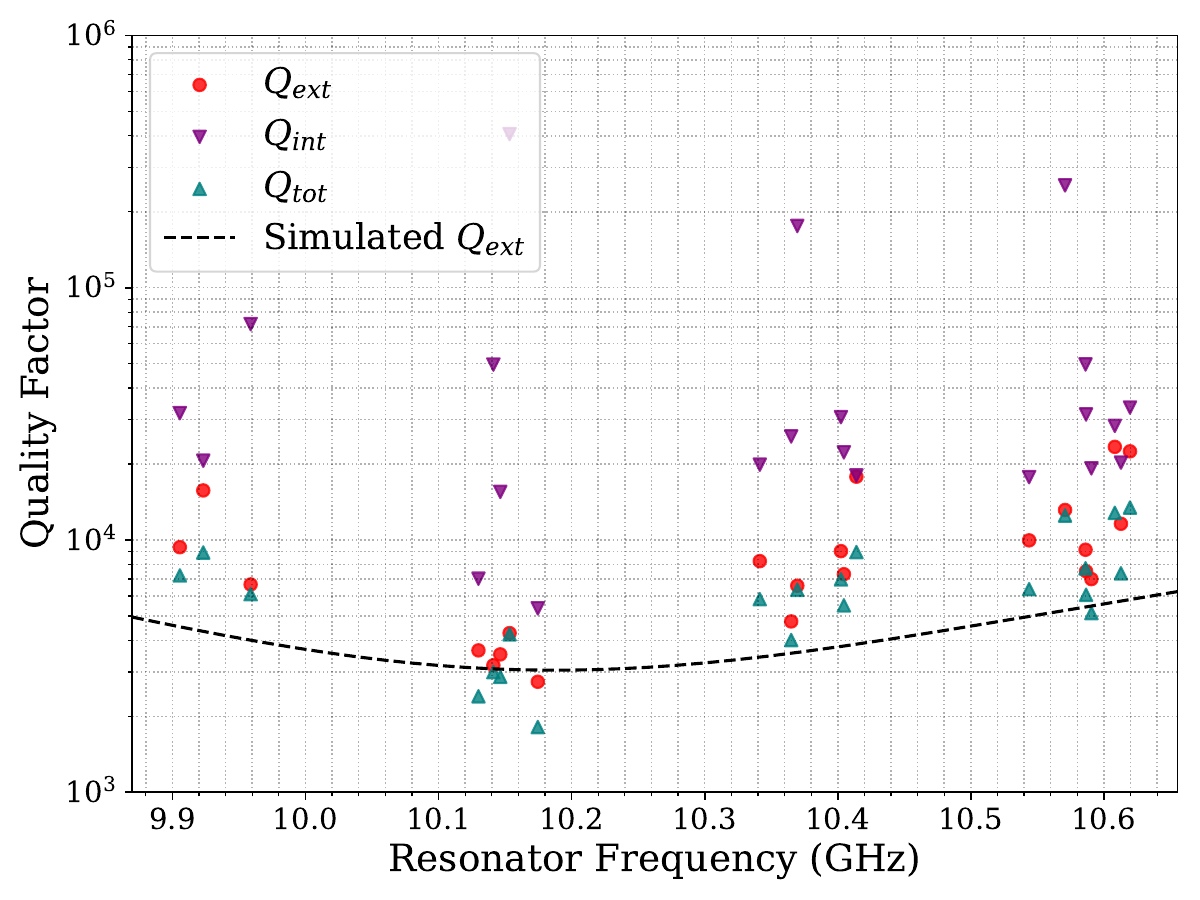}
        \put(-240,190){\small\textbf{(a)}}
        \label{fig5:sub1}
    \end{subfigure}
    \hfill
    \begin{subfigure}[b]{0.48\textwidth}
        \centering
        \phantomcaption
        \includegraphics[width=\textwidth]{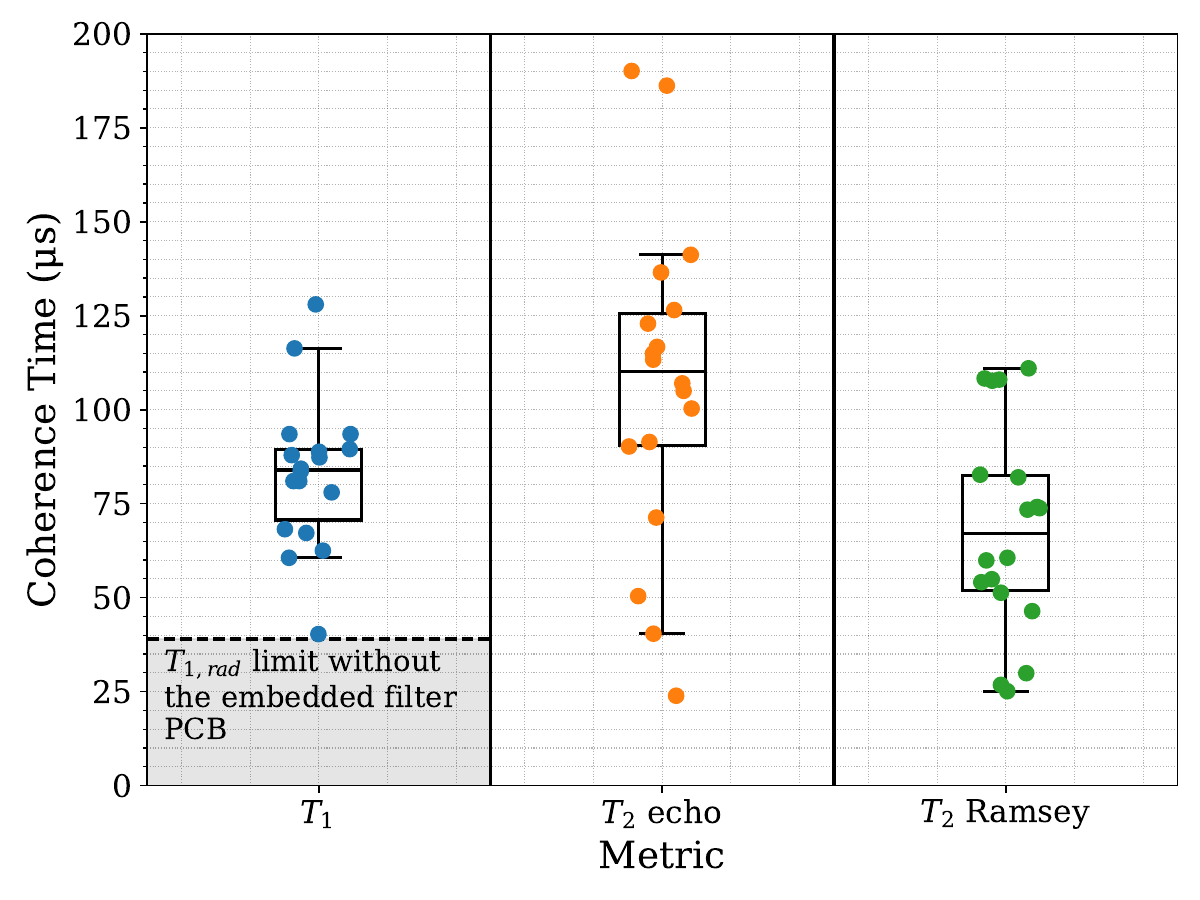}
        \put(-240,190){\small\textbf{(b)}}
        \label{fig5:sub2}
    \end{subfigure}
    \caption{\textbf{QPU metrics with embedded filters}. \textbf{a.} Quality factor measurements of the readout resonators estimated from continuous-wave spectroscopy data. The external (red \textcolor{seabornRed}{\scalebox{1.1}{$\bullet$}}), internal (purple \textcolor{seabornPurple}{\scalebox{0.9}{$\blacktriangledown$}}) and total (teal \textcolor{seabornTeal}{\scalebox{0.9}{$\blacktriangle$}}) quality factors were determined. All resonators were over-coupled by design. The quality factors data follows a trend with a minimum value at around 10.2 GHz. The shape is qualitatively consistent with the simulations; although the center frequency appears shifted by roughly 350 MHz compared to the simulations. \textbf{b.} Aggregated coherence data of 18 qubits measured on the device. Each data point represents one coherence metric ($T_1$ in blue \textcolor{mplC0}{\scalebox{1.1}{$\bullet$}}, $T_{2,Ramsey}$ in green \textcolor{mplC2}{\scalebox{1.1}{$\bullet$}}, and $T_{2,echo}$ in orange \textcolor{mplC1}{\scalebox{1.1}{$\bullet$}}) for one qubit, averaged over $\geq250$ measurement shots. The shaded region, bounded by the black dashed line at $\sim39$ \textmu s, marks the region of $T_1$'s achievable, based on our simulations, without a Purcell filter. Note how all measured $T_1$ values fall above this region.}
    \label{fig5}
\end{figure*}

%CW analysis and results
The resonators in the QPU were characterized by means of continuous-wave reflection spectroscopy; both internal and external quality factors of each resonator were determined by analyzing the spectroscopic data, using amplitude-phase fitting procedures~\cite{krkotic_algorithm_2021,wang2021cryogenic}. 21 out of the total 35 resonators were analyzed: one resonator did not produce any signal, which was attributed to a fabrication defect, and thirteen did not pass automated fitting procedures, which was attributed to in-line phase dispersion not encompassed by the fitting model. The aggregated data is shown in figure \ref{fig5}a. The estimated median internal, external and total quality factors, with their standard deviations, were $\bar{Q}_{int}=(30\pm90)\times10^{3}$, $\bar{Q}_{ext}=(8\pm6)\times10^{3}$ and $\bar{Q}_{tot}=(1/\bar{Q}_{int}+1/\bar{Q}_{ext})^{-1}=(6\pm3)\times10^{3}$ respectively. The large spread on the internal quality factors is attributed  to the uncertainty in the fitting procedures at large $Q_{int}$/$Q_{ext}$ values\cite{rieger2023fano}. Variations in the $Q_{ext}$ and the $Q_{tot}$ with frequency follow the expected dependence on the frequency profile of the embedded filter, as can be seen in fig. \ref{fig5}a. The $Q_{ext}$ spread of $\sim6\times10^{3}$ over a frequency range of $\sim700$ MHz shows the large bandwidth of the filter. In order to improve agreement with the experimental data, the simulation result shown in figure \ref{fig5}a (black dashed line) was shifted (toward higher frequencies) by +340 MHz. We attribute this shift to changes in the PCB material properties between room temperature and 11 mK.

%TD analysis and results
Time-domain measurements of the qubits were performed using custom control hardware for pulse-level microwave operation of each qubit. The microwave control signals were delivered to the qubits via the multiplexer ports, after being combined at room temperature with the readout signals used for the heterodyne measurement of the resonators field, resulting in a dispersive circuit QED measurement of the qubits state.~\cite{acharya_integration_2024, spring_high_2022}.
The coherence properties, given in fig. \ref{fig5}b and measured across 18 qubits, were determined as a median energy relaxation time $T_1 \simeq 84$ \textmu s (19 \textmu s standard deviation, 128 \textmu s best), a median Ramsey coherence time $T_{2,Ramsey} \simeq 67$ \textmu s (27 \textmu s standard deviation, 111 \textmu s best) and median Hahn echo coherence time $T_{2,echo} \simeq 110$ \textmu s (42 \textmu s standard deviation, 190 \textmu s best). To make the estimates statistically significant, each qubit coherence metric was measured at least 250 times over the course of $\sim8$ hours; then each coherence metric was averaged over these $\sim250$ samples, and finally the median over the 18 qubits was determined.

With these experiments, the readout of at least eight multiplexed signals from a single output line is demonstrated. While each embedded filter itself can support nine signal ports, in this case, one port was left unconnected to satisfy the tiling construction of the multiplexer to match the qubit layout of the processor.

%Considering that, for a qubit fundamental transition frequency $\omega_{01}$, the radiatively-limited $T_1$ time can be defined in terms of its external quality factor as $T_{1,rad}=Q_{ext}/\omega_{01}$, one can express simulated qubit mode quality factors in terms of a radiative $T_1$ limit. 
Using the average fundamental transition frequency over the qubit sample $\bar{\omega}_{01}/(2\pi)=4.1$ GHz, and the simulated qubit mode quality factors $Q_{qb,ext}(\omega)= Q_{qb}(\omega)$, one finds the radiatively-limited $T_1$ time $T_{1,rad}=Q_{qb}(\bar{\omega}_{01})/\bar{\omega}_{01}\simeq10^6/(2\pi\cdot4.1\cdot10^9~\textrm{Hz})\simeq39$ \textmu s in the case without PCB-based Purcell filter and $T_{1,rad}\simeq5.8\cdot10^8/(2\pi\cdot4.1\cdot10^9~\textrm{Hz})\simeq22$ ms in the case with PCB-based Purcell filter. All the qubits $T_1$ measurements in our statistical sample lie above the no-filter radiative $T_1$ limit, which is experimental evidence that the PCB filter is, in fact, offering Purcell protection (the radiative limit with filter is not achieved by a large margin, but this can easily be explained with other sources of loss limiting the qubits relaxation). 

Another indication of the validity of this conclusion comes from the comparison with  previous OQC-Toshiko QPU demonstrations \cite{acharya_integration_2024, 2402.17395v2}. In the current work, similar $T_1$ metrics were achieved, but with stronger external coupling (lower $Q_{ext}$) of the readout resonators.

\section{Conclusions}
In this paper, a novel 3D-integrated embedded Purcell filter, with the added functionality of multiplexing, is designed and presented. The target application is the suppression of losses induced by the Purcell effect in superconducting quantum circuits.

The integrated filter in this work limits the rate at which photons at the qubit transition frequency $\omega_{01}$ can be emitted into the resonator readout lines by introducing a microwave passband centered at 9.8 GHz with a FWHM of 0.9 GHz. This has been achieved by employing a novel method of embedding the filter within a multiplexing circuit, using conventional PCB-based multilayer technology and an antenna-like patch design. Due to it being a multilayer design, the embedded filter is shielded from both sides; this contributes to a low level of crosstalk between different multiplexing filters. Furthermore, due to the simplistic nature of the 3D layout, the entire RF circuitry of each individual filter and multiplexer is contained within a single embedded unit cell. This directly helps the design to be easily tileable and scalable for processors with large qubit counts.

Detailed finite-element electromagnetic simulations of the PCB filter, in association with a 3D-integrated QPU based on double-sided fabrication, were used to determine a 1000-fold reduction in the radiative decay rates of qubits at $\sim4.4$ GHz, relative to a readout architecture without Purcell filter; this with only a minimal impact on the external coupling rate $\kappa_{res}\simeq \omega_{res}/Q_{ext}$ of the readout resonators, allowing for fast circuit QED readout.

Experimental tests of a tiled 35--6 multiplexed filter with an OQC-Toshiko QPU allowed us to demonstrate the multiplexing capability of the filter PCB, with multiple readout resonators being interrogated simultaneously. Additionally, time-domain characterization of the qubit coherence metrics proved the compatibility of the PCB filter with the coherent operation of the QPU. Combined with the estimates of the qubit radiative decay rates, extracted from the electromagnetic simulations, the experimental qubits $T_1$ times distribution proved that the PCB filter is indeed effective in providing Purcell protection. Additional work building from these results includes the analysis of PCB materials effects on qubit loss and on readout fidelity, and the demonstration of tiled filters to support 500+ qubit devices~\cite{kennedy2026designoperationwaferscalepackages}.

\begin{acknowledgments}
The authors wish to acknowledge the support of the OQC team in this work. In particular, the authors thank the nanofabrication engineering team for fabricating the QPU chip and assisting with its installation, the mechanical design engineering team for designing and building the hardware parts necessary for the experiments, Archie McDavitt for generating the mechanical CAD graphics included in this paper, and Oscar W. Kennedy and Connor D. Shelly for careful review of this manuscript.
\end{acknowledgments}

\bibliography{_bib}

\end{document}